\documentclass[pdflatex,sn-nature]{sn-jnl}
\usepackage{graphicx}
\usepackage{amsmath,amssymb}
\usepackage{xcolor}
\usepackage{textcomp}
\usepackage{booktabs,longtable,array,calc}
\usepackage{xurl}
\usepackage{tabularx}

\usepackage[T1]{fontenc}
\usepackage{newunicodechar}
\newunicodechar{−}{\ensuremath{-}}
\newunicodechar{μ}{\ensuremath{\mu}}
\newunicodechar{ρ}{\ensuremath{\rho}}
\newunicodechar{β}{\ensuremath{\beta}}
\newunicodechar{Δ}{\ensuremath{\Delta}}
\newunicodechar{Σ}{\ensuremath{\Sigma}}
\newunicodechar{≈}{\ensuremath{\approx}}
\newunicodechar{×}{\ensuremath{\times}}
\newunicodechar{√}{\ensuremath{\surd}}
\newunicodechar{·}{\ensuremath{\cdot}}
\newunicodechar{±}{\ensuremath{\pm}}
\newunicodechar{→}{\ensuremath{\rightarrow}}
\newunicodechar{½}{\ensuremath{\tfrac12}}
\newunicodechar{≥}{\ensuremath{\geq}}
\newunicodechar{≤}{\ensuremath{\leq}}
\providecommand{\tightlist}{\setlength{\itemsep}{0pt}\setlength{\parskip}{0pt}}

\begin{document}
\title[Perturbations change which cells are measured]{Escaping cells dilute CRISPR screen responses unequally across cell types}
\author*[1]{\fnm{Xuanrui} \sur{Huang}}\email{huangxuanrui@u.nus.edu}
\affil*[1]{\orgname{National University of Singapore}, \orgaddress{\country{Singapore}}}
\abstract{In single-cell CRISPR screens, each cell receives a guide RNA that knocks down one gene, and the transcriptome of the cell is then sequenced. Screens run in many cell types are used to infer how cell types differ in their response to perturbation, to train models that transfer effects between cell types and to score such models. All three uses assume that the cells carrying a guide had their target knocked down. Yet cells grow for days before sequencing. When an essential gene is knocked down, knocked-down cells are depleted or stop dividing, whereas cells that escaped knockdown keep growing. The measured response is an average over all cells carrying the guide, so escaping cells dilute it. We define the escape fraction of a perturbation as the share of its measured cells whose target was not effectively knocked down. We estimate it by fitting the counts of the targeted gene in each cell as a mixture of knocked-down cells and cells at control level, and the estimate recovers known fractions in simulated mixtures of such cells. In most cell types the escape fraction rises with how essential the gene is, and it is higher in a later screen of the same cells, reaching about half of the measured cells in human embryonic stem cells (hESCs). The escape fraction predicts how much each response is diluted: the response size, the squared size of the reproducible expression change, falls to (1 − \emph{e})\textsuperscript{2} of that of the knocked-down cells, consistent with the measured dilution. Because the escape fraction differs between cell types, it changes all three uses. hESCs respond like K562 cells unless knockdown depletes them, and where it does, part of their weaker response comes from dilution. The coefficient that a simple transfer model learns is lowered by the dilution of the cell type it predicts. A benchmark that takes all measured cells as ground truth rewards predictions shrunk towards no change, partly because that ground truth is itself diluted. Where cells stop dividing instead, the measured cells shift towards G1, so the response also contains the expression differences between cell-cycle stages; this changes the direction of the response as well as its size. Response sizes should be compared on knocked-down cells or corrected for the escape fraction, and where cells arrest, the cell-cycle signature should be removed first.}
\maketitle
Single-cell CRISPR screens deliver to each cell a guide RNA that knocks down one gene and then sequence the transcriptome of each cell, so that the effect of a perturbation is measured in the cells that carry its guide. They now cover cancer lines\cite{ref1,ref2}, primary cells\cite{ref3}, pluripotent stem cells\cite{ref4,ref5,ref26}, differentiated cells\cite{ref6,ref7} and further genome-scale platforms\cite{ref8}. When the same perturbations are run in several cell types, knocking down a gene changes expression more in some than in others, and this difference in response size is put to three uses. The first is biological interpretation: cell types that respond less are taken to be more resistant to perturbation, as when human embryonic stem cells (hESCs) were reported to be broadly more resistant than K562 cells\cite{ref4}, or when fewer perturbations gave a discernible response in HEK293T than in HCT116 cells\cite{ref8}. The second is training models that transfer perturbation effects between cell types, which must learn from the data how response size changes from one cell type to another\cite{ref10,ref11}. The third is evaluating such models, since benchmarks score predictions against measured responses and set from them the best score a model could reach\cite{ref12,ref13}.

All three uses assume that the cells carrying a guide had their target knocked down, and the experiment does not guarantee this. Guides are delivered to a large pool of cells, which then grow for several days to two weeks before they are sequenced, and in that time a perturbation changes growth as well as expression. When an essential gene is knocked down, knocked-down cells are depleted from the culture or stop dividing, whereas cells that escaped knockdown keep growing. A cell escapes knockdown when the guide acts weakly or not at all in it, or when it was assigned a guide that did not act in it: each cell's guide is inferred from the guide transcripts sequenced with it, and a cell that carries several guides, picks up guide transcripts from other cells or yields few guide reads can be assigned the wrong one. By the time of sequencing, escaping cells make up a growing part of the cells that carry the guide. The measured response is an average over all of them, and escaping cells, whose expression has barely changed, dilute it: the more escaping cells, the smaller the measured response, however strongly the knocked-down cells respond.

Parts of this problem are known. The first Perturb-seq screens modelled the fraction of effectively perturbed cells\cite{ref28}; tools such as Mixscape and ContrastiveVI+ identify cells that appear unperturbed as escaping cells and remove or model them before analysis\cite{ref15,ref27,ref39}; errors in assigning guides to cells have been modelled\cite{ref29,ref30}; CRISPR interference effectors differ in how completely they knock down their targets\cite{ref31,ref32}; the depletion of cells has been noted as a limitation of individual screens\cite{ref25,ref26}; and signatures of cell-cycle arrest are removed from benchmark scores as systematic variation\cite{ref9}. These approaches treat escaping cells as a matter of perturbation efficiency within one dataset. Whether the fraction of escaping cells differs between cell types, how much it dilutes a measured response and whether it changes the conclusions of the three uses above have not been measured.

We examined published screens in 13 cell types and states. In the seven with enough cells per perturbation, we estimated the escape fraction of each perturbation, the share of its measured cells whose target was not effectively knocked down, by fitting the counts of the targeted gene in each cell as a mixture of knocked-down cells and cells at control level. We checked the estimate in three ways: it recovers known fractions in simulated mixtures of knocked-down and control-level cells, it does not find two groups when knockdown is uniform, and it predicts the dilution measured on independent replicates. The estimate uses only the targeted gene, which is left out of the response. A cell whose target was knocked down but whose other genes changed little therefore still counts as knocked down; classifying cells by their response instead would count it as escaping, and a weak response could then not be told apart from a cell that escaped knockdown. The escape fraction rises with how essential the gene is, is higher in a later screen of the same cells and differs between cell types. It predicts how much each response is diluted: the mean expression change of all cells is (1 − \emph{e}) times that of the knocked-down cells, and the response size, the squared size of the reproducible change, is (1 − \emph{e})\textsuperscript{2} times theirs, consistent with what we measure. Because the escape fraction differs between cell types, it changes the conclusions of all three uses. hESCs respond as strongly as K562 cells to genes whose knockdown does not deplete them, and for genes whose knockdown does, part of their weaker response comes from dilution. The coefficient that a simple transfer model learns is lowered by the dilution of the cell type it predicts. A benchmark scored against all cells rewards predictions shrunk towards no change, partly because its ground truth is itself diluted. Escaping cells change the size of responses but, in hESCs compared with K562 cells, largely leave their direction intact. Where cells stop dividing instead, the measured cells shift towards G1, so the response also contains the expression differences between cell-cycle stages, which alters its direction as well as its size. Responses compared between cell types should therefore be measured on knocked-down cells or corrected for the escape fraction.

\section{Results}

\subsection{Essential-gene knockdown leaves different cells to be measured in different cell types}

We placed 13 cell types and states on a common scale, comparing them at matched dependency, defined outside the data by the median DepMap gene effect\cite{ref19}; the most essential genes have a gene effect below −1 (Supplementary Table S1). For each perturbation we measured three things: the number of cells relative to perturbations of non-essential genes, the remaining expression of the target in the cells that are left (their counts of the targeted gene over the counts expected from control cells), and the change in the fraction of cells in G1. Remaining target expression agrees with the knockdown reported for the K562 and RPE1 screens\cite{ref1} (Spearman 0.97 and 0.96).

Both outcomes of knocking down an essential gene (Fig. 1a) appear, and which one dominates depends on the cell type (Fig. 1b). They are not exclusive, because arrested cells also stop expanding; what separates them is whether the knocked-down cells remain to be measured. In RPE1 and HepG2 cells, the cells of the most essential perturbations shift towards G1 while keeping their target well knocked down: they stop dividing and remain. In hESCs and primary CD4+ T cells, cell numbers fall and the cells that are left keep much of their target's expression (0.59 and 0.78 of control), so the measured cells are weighted towards those in which knockdown failed. K562 and Jurkat cells show neither in these aggregate measures, which understate depletion where libraries were balanced against dropout; escaping cells are nonetheless common in both (Fig. 2b). In the hESC screen the depletion is hidden by the library, which over-represents guides expected to drop out; corrected for library input, 2.5\% of the cells expected from non-targeting elements remain for the most essential genes, not the 31\% that raw counts suggest (Supplementary Fig. S1).

\begin{figure}[tbp]
\centering
\makebox[\textwidth][c]{\includegraphics{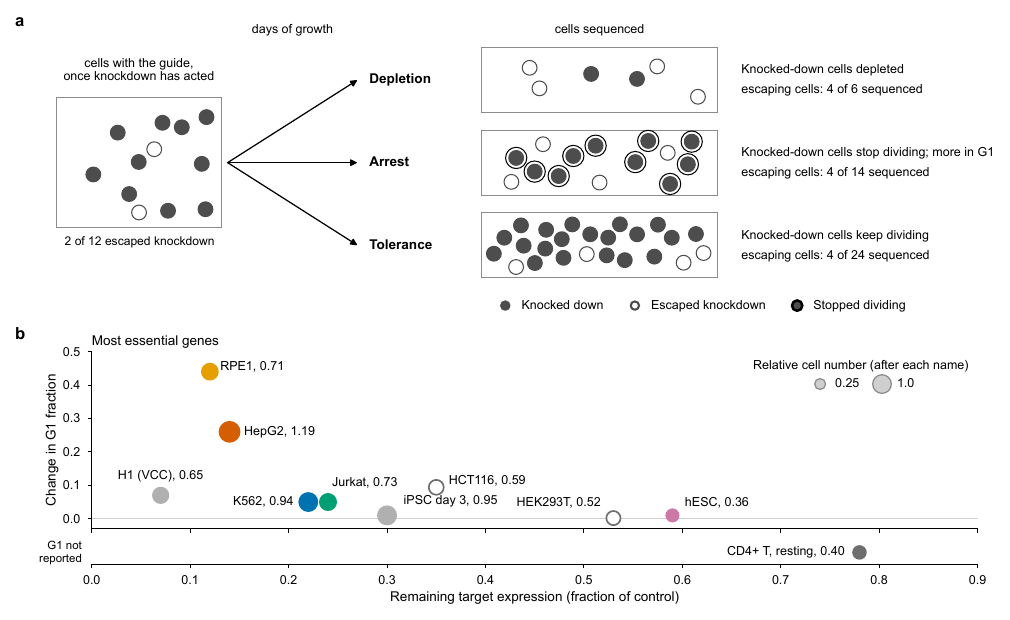}}
\caption{\textbf{After an essential gene is knocked down, cells are depleted or stop dividing, and which happens depends on the cell type.} \textbf{a}, Schematic. Cells carrying a guide (left; here 2 of 12 escaped knockdown) grow for days before sequencing. For an essential gene, knocked-down cells are depleted (depletion), stop dividing, so that more of the measured cells are in G1 (arrest), or keep dividing (tolerance). Ringed cells have stopped dividing. Escaping cells keep dividing in every case and double from 2 to 4 in each row; what differs is how many knocked-down cells remain (20 when they keep dividing, 10 when they arrest, 2 when they are depleted), so the share of escaping cells among the cells sequenced is unchanged with tolerance and rises most with depletion (right). The outcomes are not exclusive: arrested cells also stop expanding, which raises the share of escaping cells. \textbf{b}, Remaining target expression of the cells that are left (their counts of the targeted gene over the counts expected from control cells) against the change in their G1 fraction, for the most essential genes (DepMap gene effect below −1), medians over genes; point area and the number after each name, cell number relative to perturbations of non-essential genes, not corrected for library input (for hESCs, whose library over-represents guides expected to drop out, 2.5\% of the expected cells remain after correction); colours, cell types; open grey circles, HCT116 and HEK293T from the X-Atlas/Orion genome-scale screen\cite{ref8}; grey filled circles, H1 hESCs of the Virtual Cell Challenge (VCC)\textsuperscript{12} and iPSCs at day 3\cite{ref5}. The G1 change is the difference in the fraction of cells in G1 from perturbations of non-essential genes. Relative cell numbers above 1 occur where libraries were balanced against dropout, so cell number understates depletion there. n = 403 (RPE1), 694 (HepG2), 737 (HCT116), 673 (K562), 719 (Jurkat), 733 (HEK293T), 738 (hESC), 140 (iPSC), 9 (H1) and 136 (resting CD4+ T) genes. CD4+ T cells (G1 change not reported) are drawn in a separate strip; the 16 knockout lines (no target counts) and astrocytes (no G1 change) are not shown. High remaining target expression alone does not separate depletion from a weaker knockdown, as in HEK293T; together with a low cell number, as in hESCs, it points to depletion, and the per-cell analysis of Fig. 2 separates the two.}\label{fig1}
\end{figure}

\begin{figure}[tbp]
\centering
\makebox[\textwidth][c]{\includegraphics{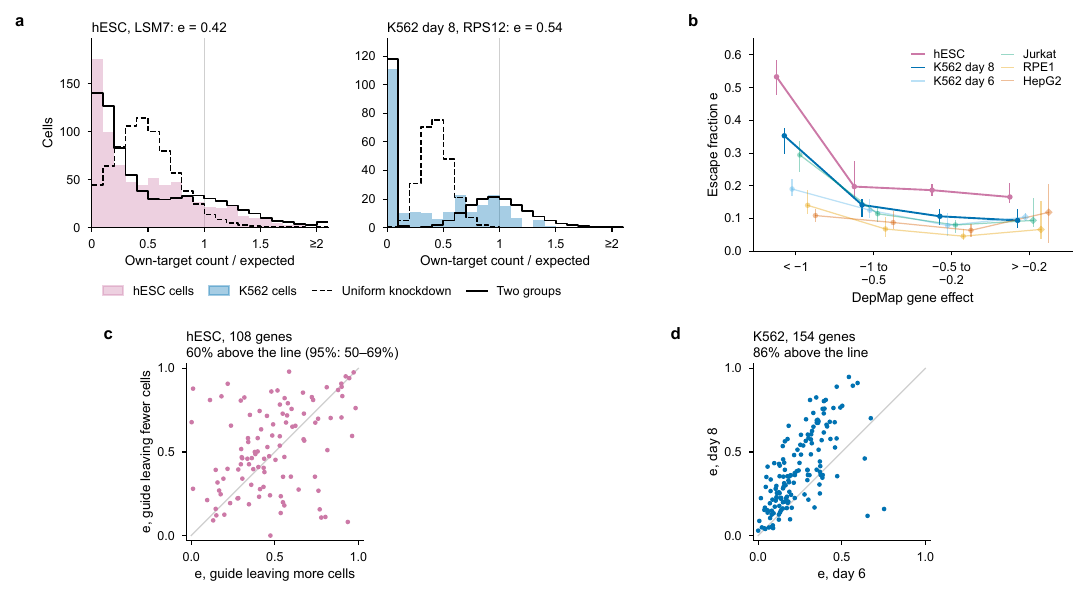}}
\caption{\textbf{The share of escaping cells can be estimated per perturbation and rises with essentiality and in a later screen.} \textbf{a}, Own-target count of each cell over its expectation from control cells for two of the most essential genes, chosen for similar mean remaining expression rather than as typical values (hESC LSM7, n = 749 cells; K562 day 8 RPS12, n = 261 cells); grey line, control level; with the histograms expected under uniform knockdown and under the two-group fit; \emph{e}, escape fraction. The two groups overlap at low counts, so the fit, not the shape of the histogram alone, separates them. \textbf{b}, Median escape fraction by stratum in five cell types (six screens); error bars, 95\% bootstrap intervals over genes; hESC and K562 day 8 drawn in full, others lighter; diamonds, strata with fewer than 20 genes; points offset sideways for legibility. n genes per stratum, most to least essential: hESC 173, 73, 62, 44; K562 day 8 162, 72, 83, 95; K562 day 6 212, 81, 35, 3; Jurkat 170, 86, 44, 10; RPE1 137, 58, 38, 8; HepG2 182, 76, 44, 11; resting CD4+ T cells are in Supplementary Table S2. \textbf{c}, hESC, most essential genes: escape fraction of the guide that left fewer cells against that of the other guide against the same gene, one point per gene (n = 108); grey line, identity; points above it are genes where the guide that left fewer cells has the higher escape fraction (share and 95\% binomial interval above the panel). \textbf{d}, K562, most essential genes: escape fraction at day 8 against day 6 (n = 154 genes fitted in both screens); grey line, identity; share of genes above it, above the panel.}\label{fig2}
\end{figure}

\subsection{The share of escaping cells can be estimated for each perturbation with enough cells}

Remaining target expression averages over cells. To see whether the cells that are left form two groups, knocked down and escaping, or share one partial knockdown, we fitted each perturbation's own-target counts with both models (Fig. 2a). Each cell's expected count at control level is the control rate of the gene times that cell's total counts. The escape fraction \emph{e} is the weight of the group at control level: the share of the perturbation's measured cells whose target was not effectively knocked down. Escaping cells thus include cells in which the guide acted weakly or not at all, cells assigned a guide they do not carry and, because knockdown varies between cells, part of the cells knocked down only partly. The fit needs enough cells per perturbation and enough counts of the target, which seven cell types provide. Most perturbations prefer two groups (45--97\% of perturbations by cell type and stratum), whereas control cells whose counts were thinned to a uniform knockdown almost never do (at most 2\%; Supplementary Table S2 and Supplementary Fig. S2). The estimate is accurate where it can be checked against a known answer: in control cells of which a set fraction kept their counts and the rest were thinned to 0.2 of control, the fitted \emph{e} was within 0.013 of the true fraction at the median for fractions of 0.1 to 0.7, and 90\% of estimates lay within about ±0.08 at 150 cells and ±0.04 at 400 cells (Supplementary Table S2). It is also specific to the targeted gene: other genes of similar expression, fitted in the same cells, prefer two groups for only 1--4.5\% of perturbations (Supplementary Fig. S2). The test detects knockdown that is not uniform rather than strictly two populations: control cells thinned to a knockdown that varies between cells (standard deviation 0.2) are called two groups for 78--88\% of genes, with \emph{e} of about 0.1, and when its level is fitted freely the escaping group sits at 0.6--0.8 of control except for the most essential genes in hESCs (0.89; Supplementary Table S2 and Supplementary Fig. S2).

The escape fraction rises with how essential the gene is (Fig. 2b). For the most essential genes, about half of the measured cells escaped knockdown in hESCs, a third in K562 cells at day 8 and in Jurkat cells, and a fifth or less in RPE1 and HepG2 cells and in K562 cells at day 6, against 0.07--0.17 for non-essential genes (values for every perturbation in Supplementary Data 1). The differences remain between cell types screened with the same CRISPR interference effector and library, K562 against HepG2 and RPE1 against Jurkat cells\cite{ref1,ref2}. The rise holds when the target's expression and the number of cells are controlled, and it is absent in HepG2 cells. In resting CD4+ T cells, most of the most essential perturbations keep fewer than 20 cells, so the escape fraction can be fitted only for the eight genes in each of the two most essential strata that remain; among those it does not rise with essentiality (Spearman −0.03; Supplementary Table S2), and these cells are not drawn in Fig. 2b. The baseline for non-essential genes has other sources, uneven knockdown among them, and guide calls: in hESCs, where guide counts are available for each cell, it falls to a few per cent among cells with an unambiguous guide call, whereas the rise with essentiality remains (Supplementary Table S2).

\subsection{Escaping cells are more common where more cells were depleted and at a later harvest}

If the rise comes from the depletion of knocked-down cells, perturbations that deplete more cells should have a higher escape fraction. In hESCs, where cell numbers can be corrected for the library, they do: among perturbations of matched dependency, those that retained fewer cells have a higher escape fraction (Spearman −0.42 to −0.62 within strata). In the other cell types the relation is weak, absent or reversed (Supplementary Table S2), because cell numbers there measure depletion poorly: most of these libraries were balanced against dropout, and a guide that acts weakly leaves both more cells and more escaping cells. For the most essential genes, the guide that left fewer cells has the higher escape fraction of the two guides against the same gene in 60\% of 108 genes, an interval (50--69\%) that reaches chance (Fig. 2c), although a guide that depletes more cells is, if anything, the more effective one.

Depletion takes time, so the escape fraction should be higher at a later harvest. In K562 cells, the most essential genes have a higher escape fraction at day 8 than at day 6 in 86\% of the 154 genes measured in both screens (Fig. 2d). The two screens used the same guides in the same cell line but also differ in library, sequencing depth and batch, so this comparison supports, but does not isolate, an effect of time. In one K562 screen sampled at days 4, 7, 10 and 14\cite{ref20}, perturbations that deplete cells reach their largest response at day 7 and then lose it as their remaining target expression rises, whereas active perturbations that do not deplete cells keep theirs (few targets; Supplementary Fig. S3). A comparison between cell types therefore also compares the times at which each was harvested.

\subsection{The escape fraction predicts how much each response is diluted}

If escaping cells contribute little response, the mean expression change of all cells is (1 − \emph{e}) times that of the knocked-down cells, and the response size, which multiplies the changes measured in two independent replicates, is (1 − \emph{e})\textsuperscript{2} times theirs. We tested this perturbation by perturbation, measuring dilution as the response size of all cells over that of knocked-down cells, both between independent replicates. Measured dilution follows the dilution predicted from \emph{e} with a slope close to one and no offset in hESCs and K562 cells (least-squares slopes 0.97 and 0.93; 1.17 and 1.06 after de-attenuation, which treats all disagreement between the two replicates' estimates of \emph{e} as noise and therefore gives an upper value; offsets within 0.1 log2 units; Fig. 3a, b). In RPE1 cells, where only random halves exist and the slope cannot be de-attenuated, least squares gives 0.97 and a robust fit 1.32, with an offset of −0.07 (Fig. 3c). Escaping cells therefore add little response of their own, and (1 − \emph{e})\textsuperscript{2} corrects each response for dilution by escaping cells.

\begin{figure}[tbp]
\centering
\makebox[\textwidth][c]{\includegraphics{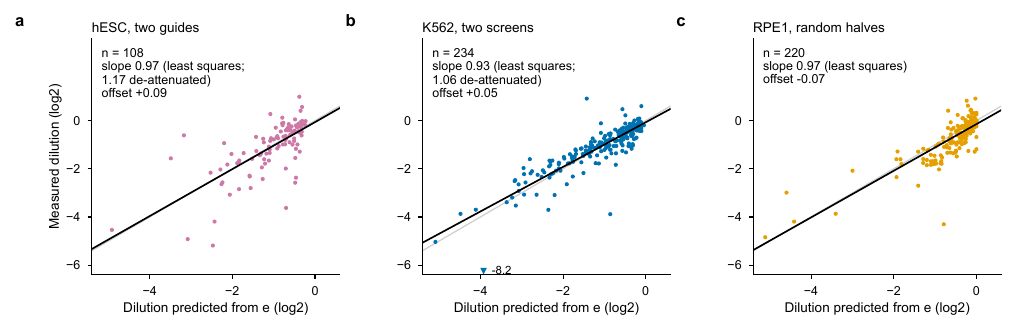}}
\caption{\textbf{The escape fraction predicts how much each response is diluted.} Measured dilution, the response size of all cells over that of knocked-down cells, against the dilution predicted from the escape fraction, (1 − \emph{e})\textsuperscript{2}, both log2, one point per gene. The response size is the product of the expression changes measured in two independent replicates, summed over genes, so it scales with the square of the reproducible change. Knocked-down cells are classified one by one from the targeted gene's counts, and the response is measured on the other genes with the targeted gene left out, so the escape fraction and the measured dilution share no data beyond the targeted gene's counts. \textbf{a}, hESC (guide halves; n = 108 genes); \textbf{b}, K562 (two independent screens; n = 234); \textbf{c}, RPE1 (random halves; n = 220). Black line, least-squares fit; in \textbf{a} and \textbf{b} the de-attenuated slope, which corrects for noise in each estimate of \emph{e} and is an upper value, is given beside it. Grey line, identity; offset, median of measured minus predicted; the one point below the axis in \textbf{b} is drawn at its edge with its value.}\label{fig3}
\end{figure}

Selecting knocked-down cells needs a matched control. Taking cells with a low count of one gene also selects a cell state: control cells taken from the low tail of a gene gave a response larger than the signal itself. A classification control, made by thinning control cells to the knockdown level and classifying them by the same rule, reproduces how strongly the classifier selects cell state; its response is 1--2\% of the signal, also when unthinned cells are mixed in at the observed escape fraction. Measured directly in hESCs, the cells classified as escaping respond at about a twentieth of the knocked-down cells of the same gene (Supplementary Table S2). Without it, dilution can be badly overstated, as in CD4+ T cells, where most of the apparent excess disappears once the control is subtracted (Supplementary Fig. S4). Where only aggregate counts exist, the remaining target expression gives a cruder estimate of \emph{e}, which predicts dilution with the same slope but misses the baseline escape fraction present for every perturbation (Supplementary Fig. S4).

\subsection{The weaker hESC response follows the depletion of cells}

hESCs were reported to respond less to perturbation than K562 cells, also when the analysis was restricted to genes knocked down efficiently in hESCs, and this was read as broad resistance of the hESC state to perturbation\cite{ref4}. We repeated the comparison with the same genes paired across the two cell types and response sizes measured between independent replicates: the two guides of each gene in hESCs and two independent K562 screens (Fig. 4a). Over all shared genes the hESC response is far weaker (ratio 0.17; Fig. 4b), reproducing the report. On genes knocked down well in both cell types (remaining target expression at most 0.3), the responses are comparable (ratio 1.07, 95\% interval 0.84--1.30; Fig. 4b), and the ratio stays between 0.66 and 1.21 over nine expression and knockdown thresholds (Supplementary Tables S3 and S4; Supplementary Fig. S5).

\begin{figure}[tbp]
\centering
\makebox[\textwidth][c]{\includegraphics{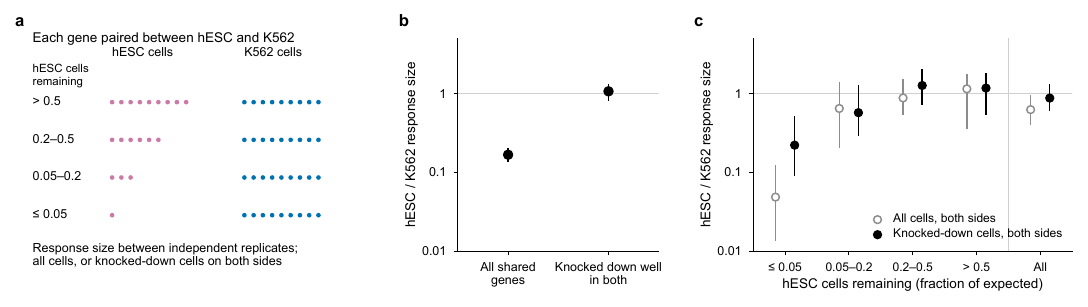}}
\caption{\textbf{Embryonic stem cells respond less than K562 cells mainly where knockdown depletes them.} \textbf{a}, The comparison: each gene paired between hESC and K562, response sizes between independent replicates, genes grouped by how many of the expected hESC cells remain. \textbf{b}, hESC-to-K562 ratio of median response sizes over the shared genes with an hESC response size between guide halves (n = 1,308 of 1,328) and over genes knocked down well in both cell types (remaining target expression at most 0.3; n = 347). \textbf{c}, The same ratio by hESC cells remaining (fraction of expected, corrected for each guide's representation in the library), on the 99 genes with enough knocked-down cells in both cell types; these genes are less depleted than those in \textbf{b}, so their all-cell ratio is higher (``All''). Shown all cells on both sides and knocked-down cells on both sides (n = 13, 24, 26 and 36 genes by bin; 99 overall). Error bars, 95\% bootstrap intervals over genes; \textbf{b} and \textbf{c} share the vertical scale.}\label{fig4}
\end{figure}

The weaker response follows the depletion of cells. Grouped by how many of the expected hESC cells remain, the ratio is near one where more than half remain and falls to a few per cent where fewer than 5\% remain (Fig. 4c). Restricting both cell types to knocked-down cells, since K562 cells also deplete knocked-down cells of essential genes, closes about half of the gap on a log scale where hESCs lose most cells (from 0.05 to 0.22 of the K562 response) and about 40\% across the most essential genes. The rest remains with knocked-down cells on both sides: the knocked-down hESCs that are left respond at a fifth to a third of their K562 counterparts. Part of this remainder reflects a shallower knockdown: where hESCs are depleted, the knocked-down hESCs that remain often keep more of their target than knocked-down K562 cells, and at similar depth the gap narrows, although it does not close for the most depleted genes (Supplementary Table S2). Whether the knocked-down hESCs that remain resist perturbation or are what remains after the most responsive cells were eliminated cannot be told apart, because eliminated cells are not measured; hESCs are known to eliminate cells under stress rapidly\cite{ref33,ref34,ref35}. The weaker response of hESCs is therefore absent for genes knocked down well in both cell types and for genes whose knockdown leaves most hESCs. Where knockdown depletes hESCs, it combines dilution by escaping cells with a weaker response of the knocked-down cells that remain, part of which follows a shallower knockdown.

A second published difference is compatible with the same reading but is not resolved here. The weaker response of HEK293T than HCT116 cells to the most essential perturbations\cite{ref8} goes with a uniformly higher remaining target expression in HEK293T and is no longer detectable among genes whose remaining expression is equal in the two lines, but the matching premise was not met (remaining expression reproducible between halves at Spearman 0.45, below the 0.5 required) and the shift is not significant (Supplementary Fig. S6).

\subsection{Escaping cells change what a transfer model learns and how it is scored}

A model that transfers responses between cell types must learn how much larger or smaller they are in the target cell type. The simplest such model multiplies the response in one cell type by a single coefficient to predict the response in another, choosing the coefficient that minimises the prediction error. Predicting hESCs from K562 cells for the most essential genes, the coefficient learned from all hESC cells is 0.27, and from knocked-down hESC cells 0.45: the coefficient learned from all cells is 1.65 times smaller (95\% interval 1.53--1.79), and 1.6--1.7 times with Jurkat, HepG2 or RPE1 cells as the source (Fig. 5a and Supplementary Table S5). This factor measures the dilution of the hESC ground truth alone; the source cell type is diluted too, which raises the coefficient, so the net effect of the difference between the two escape fractions is smaller. Whether more complex models learn the same was not tested.

\begin{figure}[tbp]
\centering
\makebox[\textwidth][c]{\includegraphics{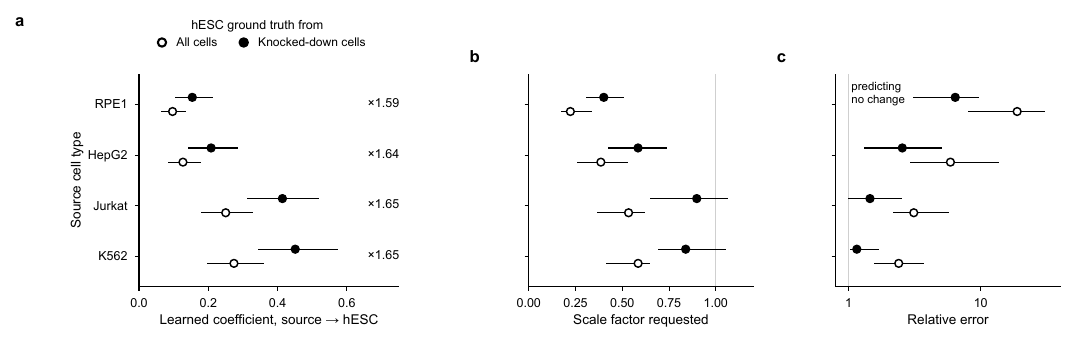}}
\caption{\textbf{Escaping cells lower the coefficient that a single-coefficient transfer model learns and change a benchmark's scores.} Most essential genes, predicting hESC from each of four source cell types on shared genes (n = 62 K562, 54 Jurkat, 54 HepG2, 56 RPE1). \textbf{a}, Coefficient learned by a single-coefficient transfer model, which predicts the hESC response as the coefficient times the source cell type's response, with hESC ground truth from all cells (open) or from knocked-down cells (filled); right, ratio of the two coefficients. \textbf{b}, The benchmark's scale factor, the size of the ground truth over that of the prediction, below 1 when a smaller prediction is asked for, with the same two ground truths. \textbf{c}, The benchmark's relative error (1, the error of predicting no change; log scale). Bars, 95\% bootstrap intervals over genes. Ground truths from the two guide halves of hESC; the source's response size from random halves of each source screen, the same for both ground truths.}\label{fig5}
\end{figure}

Benchmark scores change accordingly, partly because a ground truth measured on all cells is itself diluted. We score predictions by their squared error relative to that of predicting no change, an error measure of the kind current benchmarks use\cite{ref12}, and by the scale factor, the size of the ground truth over that of the prediction, which asks for a smaller prediction when it is below one. K562 responses taken as predictions for hESCs are asked to be shrunk to about 0.6 of their size and are rated more than twice as bad as predicting no change; with hESC ground truth restricted to knocked-down cells, the requested shrinkage largely disappears and the error falls by half, although it stays slightly above that of predicting no change (Fig. 5b, c). The shift has the same direction with Jurkat, HepG2 or RPE1 cells as the source. Ground truth from a cell type that is depleted is also mostly noise for its most essential genes, because few cells remain for each (Supplementary Fig. S7). In a genome-scale library only a few per cent of perturbations are heavily diluted, but in libraries built on essential genes, as many benchmarks and comparisons between cell types are, a substantial fraction are (Supplementary Fig. S7).

\subsection{Escaping cells change the size of responses and largely leave their direction intact; arrest changes both}

If escaping cells add little response, restricting to knocked-down cells should change how strongly a gene responds but not which genes respond. We measure agreement of direction between two cell types as the correlation of their expression changes over genes, corrected for replicate noise. In hESCs it does not change the direction of the response relative to K562 cells for the most essential genes: the change in agreement is −0.003, and a decrease of more than 0.025 can be excluded (Fig. 6a); with the classification control subtracted, the change is −0.007 and a decrease of more than 0.035 can be excluded (Supplementary Table S2). For less essential genes agreement is slightly lower (by about 0.03, and about 0.05 with the classification control subtracted; Supplementary Table S2), for reasons we have not identified. In this comparison, which genes respond is therefore largely unaffected by escaping cells when agreement is corrected for noise; measures that are not, such as overlaps of differentially expressed genes, still lose signal as responses are diluted.

\begin{figure}[tbp]
\centering
\makebox[\textwidth][c]{\includegraphics{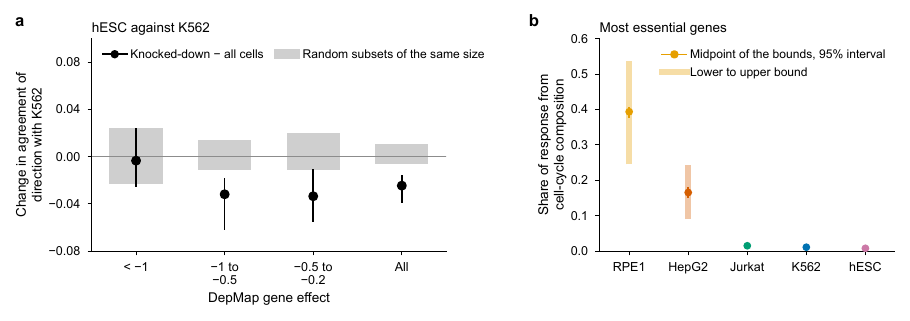}}
\caption{\textbf{Escaping cells change the size of responses and largely leave their direction intact; arrest adds a cell-cycle component to the response.} \textbf{a}, Change in agreement of direction with K562 (correlation of expression changes over genes, corrected for replicate noise; 0.58 with all hESC cells in the most essential stratum) when hESC is restricted to knocked-down cells, by stratum and overall (the least essential stratum has too few genes): median over genes with 95\% bootstrap interval (n = 41, 41, 20; 104 overall); grey band, 2.5th to 97.5th percentile of the same median for 200 size-matched random subsets of all cells. \textbf{b}, Share of the response of the most essential genes carried by the cell-cycle composition term, the expression change that the shift in cell-cycle stages alone would produce (control expression in each stage, weighted by the perturbed cells' stage proportions). Because composition and within-stage change overlap, the share has a lower and an upper bound (bar); dot, their midpoint, median over genes with 95\% bootstrap interval; n = 661 (RPE1), 659 (HepG2), 665 (Jurkat), 756 (K562) and 594 (hESC) genes. How removing this component changes direction is shown in Supplementary Fig. S8.}\label{fig6}
\end{figure}

The other outcome, arrest, changes direction as well as size. Where perturbed cells stop dividing, the measured cells shift towards G1, and because cell-cycle stages differ in expression, this change in composition alone produces expression changes, which we call the composition term. The composition term carries between 0.24 and 0.54 of the response of the most essential genes in RPE1 cells (midpoint 0.39) and between 0.09 and 0.24 in HepG2 cells (midpoint 0.17), against 0.008--0.015 in Jurkat, K562 and hESCs (Fig. 6b, Supplementary Table S6 and Supplementary Fig. S8). Removing it changes agreement with K562 cells, mostly by removing an arrest signature shared across genes (Supplementary Fig. S8), and it accounts for part of the published similarity of RPE1 and HepG2 among four cancer lines screened with one library\cite{ref2} (Supplementary Fig. S6). The same calculation with a pluripotency score in place of cell-cycle stage gives a share of 0.003 in hESCs: no shift in differentiation state is detected with this marker score.

\begin{center}\fbox{\begin{minipage}{0.94\linewidth}\small
\textbf{Box 1 \textbar{} Comparing perturbation responses between cell types}

\begin{enumerate}
\def\labelenumi{\arabic{enumi}.}
\tightlist
\item
  Report, for each perturbation, the escape fraction from a per-cell fit, together with its cell number relative to non-targeting controls, normalised for library input where the library was balanced against dropout.
\item
  To compare response size, use knocked-down cells on both sides, with a classification control of matched strength, or, for groups of perturbations, divide response sizes by (1 − \emph{e})\textsuperscript{2} (for a single perturbation this correction is too noisy). Comparisons of direction are largely unaffected by escaping cells when agreement is corrected for noise.
\item
  Compare cell types at matched time after knockdown has taken effect, not at matched harvest day alone.
\item
  Where perturbed cells arrest, remove the cell-cycle composition term before comparing response size or direction.
\end{enumerate}
\end{minipage}}\end{center}

\section{Discussion}

A measured perturbation response is an average over cells that the perturbation itself has selected. When an essential gene is knocked down, the cells in which knockdown succeeded are depleted or stop dividing, and the cells that are measured become, to a degree that differs with cell type and time of harvest, those that escaped. The escape fraction can be estimated for each perturbation with enough cells from the targeted gene's own transcripts, and it predicts how much each response is diluted. Because it differs between cell types, it changes what comparisons between cell types conclude about how strongly cells respond, while leaving largely intact what they conclude about which genes respond; arrest, the other outcome, changes both.

Unperturbed cells are not a new observation, and tools exist to remove them within a dataset\cite{ref15,ref27,ref39}. What changes when cell types are compared is that the escape fraction is not a nuisance constant: it rises with how essential the gene is and was higher in a later screen, so it differs systematically between the two sides of a comparison. Classifying cells by the targeted gene's own transcripts, rather than by their response, keeps the estimate independent of the response being compared, and a matched classification control keeps it honest.

The three uses each have a direct remedy. Biological claims that one cell type resists perturbation can be checked on knocked-down cells and on genes whose knockdown depletes neither cell type. Models that transfer effects between cell types can be trained on knocked-down cells or on responses corrected by (1 − \emph{e})\textsuperscript{2}, and the escape fraction can be supplied as a covariate of each perturbation. Benchmarks can take their ground truth from knocked-down cells, or flag the perturbations of essential genes in cell types that are depleted. For new screens, harvesting soon after knockdown takes effect and at matched times may keep the escape fraction low in the first place, and reporting the escape fraction of each perturbation, which needs only the per-cell counts every CRISPR interference screen already has, given enough cells per perturbation, lets others correct for it.

Our account stops where the data stop. The cells that a perturbation eliminated cannot be measured, so whether they would have responded more is unknown; in hESCs the remaining cells span a continuum from control level to shallow knockdown, which suggests that cells are depleted according to how deeply they knocked down the target. Within a stratum of matched dependency, how many cells remain differs between perturbations far beyond its error, and we could not predict it from guide dose or from programmes induced in the remaining cells (Supplementary Fig. S9). The analyses cover cells in culture, mostly dividing, perturbed by CRISPR interference or knockout, not activation, drug, in vivo or combinatorial screens, and only the simplest transfer model was tested. Each cell type comes from one study, so cell type cannot be fully separated from protocol, although the differences persist between cell types that share effector and library. Independent replicates exist for only part of the data, so cell types with random halves only are compared as rankings, and the symmetric hESC--K562 comparison rests on the 99 genes classifiable in both. No dataset here reads out the target protein in each cell, so the escape fraction is validated against measured dilution and simulated mixtures rather than an independent measurement of knockdown, and ambient RNA, not modelled here\cite{ref37,ref38}, adds a floor to remaining target expression that affects comparisons of its absolute level between cell types. A direct test would manipulate what happens to perturbed cells rather than observe it: preventing their depletion in a cell type that loses them should restore the response to the most essential perturbations and change nothing in a cell type that tolerates them.

\section{Methods}

\subsection{Analysis rules applied throughout}

\begin{enumerate}
\def\labelenumi{\arabic{enumi}.}
\tightlist
\item
  A response is called reproducible only between independent replicates (two guides of a gene, two independent screens, donors, replicate plates), or else accompanied by a null-perturbation control; random splits of one perturbation's cells share clone- and well-level state and overstate reproducibility\cite{ref16}.
\item
  Every control is designed so that it can fail.
\item
  Quantities computed from noisy estimates (benchmark ground truth, ratios between cell types, regression on noisy predictors) are also reported in a denoised or de-attenuated form.
\item
  Before any difference between replicates is explained, it is tested against the error within replicates; where the test does not pass, the question is reported as unanswerable in that dataset.
\item
  Cell-cycle stage calls and ΔG1 are used only where control cells are demonstrably cycling.
\item
  The reference of a null distribution is drawn from the same cells as the samples it is compared with.
\item
  Target genes retained to count their own transcripts are excluded from response vectors unless they pass the expression threshold.
\end{enumerate}

\subsection{Data and preprocessing}

All data are published (Data availability). Guide assignments were taken from the original studies wherever per-cell calls are deposited (K562, RPE1, Jurkat, HepG2, CD4+ T cells, X-Atlas/Orion, iPSC, the 16 knockout lines and the K562 time course). CRISPR interference effectors, as stated in the original studies: a KOX1 KRAB domain in K562 cells\cite{ref1} and HepG2 cells\cite{ref2}, a ZIM3 KRAB domain in RPE1 cells\cite{ref1} and Jurkat cells\cite{ref2}. For the hESC screen, which deposits guide UMI counts, a cell was assigned to its top guide when that guide had at least 3 UMIs and at least 80\% of the cell's guide UMIs; other cells were not used. Non-targeting elements, as labelled in each study, define the controls. Library size is the cell's total gene-expression UMIs. Guide parity is the rank of the called guide's name among the gene's guides, modulo 2, so the two guide halves of a gene are disjoint sets of transductions; random halves split cells by a hash of the cell barcode. Supplementary Table S7 gives, for each analysis, the rule that selects its genes and the number that enter.

\subsection{Essentiality strata}

Dependency was defined as the median gene effect across cell lines in DepMap 24Q4, in four strata: below −1, −1 to −0.5, −0.5 to −0.2, above −0.2. All comparisons between cell types are made within a stratum.

\subsection{Outcomes of perturbation}

\emph{Relative cell number} is the median cell count divided by the median of the non-essential stratum of that cell type; \emph{remaining target expression} is the target gene's raw count summed over the perturbation's cells divided by the control rate of that gene times the summed library size, estimated when the expected count exceeds 20; \emph{ΔG1} is the G1 fraction of the perturbation's cells minus that of control cells. Stage was called per cell from the S and G2/M gene sets of Tirosh et al.\cite{ref23}, each scored against ten expression-matched control genes, G1 being the cells with both scores at or below zero (Supplementary Fig. S10).

\emph{Library-input normalisation.} Where the plasmid pool over-represents guides expected to drop out, normalised abundance is (cells per element / oligo copies) divided by the same ratio for non-targeting elements; the two guides of a gene were normalised separately.

\subsection{Response size and ratios}

Effects are pseudobulk log ratios: raw counts summed over a group of cells plus 0.5, divided by that group's library size, against the same quantity for control cells of the same replicate. Response genes reach 0.5 CP10K (counts per 10,000) in the cell type's controls (1 CP10K in hESC--K562 comparisons, in all three screens); the target's own gene is set to zero. Response size is the cross-signal of two independent replicates, the sum over genes of the product of their effects; it estimates the squared size of the reproducible effect and can be zero or negative when there is none. Ratios between cell types are ratios of medians over all genes of a set, with non-positive response sizes kept; the ratio of sums and the median of per-gene log ratios over genes positive on both sides are reported alongside (Supplementary Fig. S5). Reducing the pseudocount to 0.05 changed the all-gene ratio from 0.168 to 0.171, and thinning the counts of well-represented hESC perturbations to the cell numbers of the most depleted ones changed their response sizes by a factor of 0.994. If a fraction \emph{e} of a perturbation's cells is at control level, its mean expression change is (1 − \emph{e}) times that of the knocked-down cells, which on the log scale holds approximately for effects that are not large, and its response size is (1 − \emph{e})\textsuperscript{2} times theirs. Agreement between cell types is the de-attenuated correlation ρ = ½(\emph{h}\textsubscript{0}·\emph{k}\textsubscript{8} + \emph{h}\textsubscript{1}·\emph{k}\textsubscript{6}) / √((\emph{h}\textsubscript{0}·\emph{h}\textsubscript{1})(\emph{k}\textsubscript{6}·\emph{k}\textsubscript{8})) for hESC guide halves \emph{h} and the two K562 screens \emph{k}, undefined when either cross-signal is not positive.

\subsection{Per-cell classification of knocked-down and escaping cells}

For each cell, the expected count of its target gene is the control rate times that cell's library size; a Poisson log likelihood ratio between control level and a knockdown level \emph{r}, with thresholds at ± log 9, assigns the cell to knocked-down, escaping or undetermined (Supplementary Fig. S11). \emph{r} is 0.2 in hESC and, in the other cell types, the median remaining target expression of the non-essential stratum (at least 0.05). Genes are analysable when the control expectation is at least three counts. The classifier forms groups of cells; the escape fraction itself comes from the negative-binomial fit below.

\emph{Classification controls.} The thinning control thins each control cell's count of the target gene binomially to the knockdown level and classifies it by the same rule; the mixed control leaves a fraction \emph{e} (0.3 or 0.5) of control cells unthinned before classifying. The classified group's effect, minus the effect of its matched control, gives the knocked-down effect used in every comparison and dilution curve, including CD4+ T cells and the hESC ground truth of the benchmark, where the thinning control of random half \emph{u} is paired with guide half \emph{u}. The one exception is agreement of direction (Fig. 6a), computed without the control so that it can be compared with size-matched random subsets of cells; the value with the control subtracted is given alongside.

\subsection{Escape fraction: mixture versus uniform partial knockdown}

In seven cell types and states (hESC; K562 at day 6, essential-gene library, and day 8, genome-scale library; RPE1; Jurkat; HepG2; resting CD4+ T cells of donor 1), for each analysable gene with at least 30 cells, own-target counts \emph{c}\textsubscript{i} with control-level expectations \emph{μ}\textsubscript{i} were fitted by two models: uniform, \emph{c}\textsubscript{i} \textasciitilde{} NB(\emph{r} \emph{μ}\textsubscript{i}, \emph{k}); mixture, (1 − \emph{e}) NB(\emph{r}\textsubscript{kd} \emph{μ}\textsubscript{i}, \emph{k}) + \emph{e} NB(\emph{μ}\textsubscript{i}, \emph{k}), with the escaping component at control level. The size \emph{k} was estimated per gene from control cells by moments (Var = μ + μ\textsuperscript{2}/\emph{k}) and shared by both models and by the knocked-down and escaping groups of the mixture. The mixture was fitted by expectation--maximisation from three starting points (at most 300 iterations, tolerance 10\textsuperscript{−6}), with \emph{r}\textsubscript{kd} updated by bounded one-dimensional optimisation; the mixture likelihood was taken as at least the uniform one, which it nests, and the model with the lower BIC (one and two free parameters) was preferred when the difference exceeded 10. \emph{e} is reported for every fitted gene, whether or not the mixture is preferred; taking \emph{e} = 0 where it is not preferred gives dilution slopes within the intervals of the primary estimate. Stratum medians carry 95\% bootstrap intervals over genes (2,000 draws); partial Spearman correlations rank-transform \emph{e}, DepMap effect, log mean expected count and log cell number and correlate the residuals of the first two after regression on the last two. The control drew the same number of control cells as the perturbation, thinned their counts by Binomial(\emph{c}, 0.5) and fitted them the same way. A Poisson version is reported alongside.

\emph{Recovery test.} For 60 randomly chosen analysable genes per cell type (hESC, K562 day 8), \emph{n} control cells (50, 150 or 400) were drawn; a fraction \emph{e} (0, 0.1, 0.3, 0.5, 0.7 or 0.9) kept their counts and the rest were thinned by Binomial(\emph{c}, 0.2); the result was fitted as above.

\emph{Guide purity and guide pairs.} In hESC, each cell's guide purity is the UMI count of its assigned (top) guide divided by all guide UMIs in the cell, with the guide call itself unchanged. Each gene's cells were split into thirds by purity, and the mixture was fitted in each third with at least 30 cells. For guide pairs, the mixture was fitted separately to the cells of each guide parity (at least 30 each) and compared with that parity's library-normalised abundance.

\emph{Controls for the two-component test.} (i) Synthetic continuous knockdown: control cells of 40 analysable genes per cell type were thinned cell by cell with remaining expression drawn from a Beta distribution of mean 0.3 and standard deviation 0, 0.1, 0.2 or 0.3, with no cells at control level, and once more with \emph{e} = 0.3 added; (ii) non-target genes: for each analysable hESC gene, up to five other genes expressed at 0.5--2 times its own level were fitted in that perturbation's cells; (iii) free second component: the escaping level was fitted in {[}0.37, 1.8{]} of control.

\subsection{Dilution curves}

Measured dilution is log2(\emph{M}\textsubscript{all} / \emph{M}\textsubscript{kd}), both cross-signals between independent replicates (guide halves in hESC and CD4, the two screens in K562, random halves in RPE1), with the matched classification control subtracted from the knocked-down side. Predicted dilution is log2(1 − \emph{e})\textsuperscript{2}, with \emph{e} from the per-cell fit (clipped at 0.95) or, from aggregate counts, \emph{e}\textsubscript{R} = (\emph{R} − \emph{r}\textsubscript{kd})/(1 − \emph{r}\textsubscript{kd}) with \emph{R} the remaining target expression and \emph{r}\textsubscript{kd} the non-essential median. Each predictor was also computed in each of the two replicates; their correlation ρ across genes gives the reliability of the pooled value, 2ρ/(1 + ρ), and the de-attenuated slope is the least-squares slope divided by it, with a 95\% bootstrap interval over genes (1,000 draws) and a null from permuting the pooled predictor within strata (200 permutations). The offset is the median of measured minus predicted dilution. Genes enter when both response sizes are positive. The reliability treats any difference between the two replicates' estimates of \emph{e} as noise, including real differences between screens or guides, so the de-attenuated slope is an upper value and the least-squares slope a lower one; both are reported.

\subsection{The hESC--K562 comparison}

For the 1,328 genes shared by hESC and both K562 screens, the hESC library-normalised abundance was binned (≤ 0.05, 0.05--0.2, 0.2--0.5, \textgreater{} 0.5). For the symmetric comparison, knocked-down effects were formed on both sides on one gene set (1 CP10K in all three screens): in hESC per guide parity, in K562 per screen with both random halves pooled, each with its matched thinning control subtracted; response sizes are cross-signals between guide parities and between screens. Genes enter when all four response sizes (all cells and knocked-down cells, both cell types) exist. The two designs differ: hESC response sizes come from two different guides per gene, which removes effects specific to one guide, whereas the two K562 screens used the same guides, which keeps them; this lowers hESC response sizes relative to K562. The K562 all-cell response sizes from this extraction agree with those of the main comparison (Pearson 0.999). Against each K562 screen alone, random halves were used on both sides.

\subsection{Further published comparisons}

For the four cancer lines of one essential-gene library, de-attenuated correlations were computed for the six pairs on shared genes and response genes, with total effects, with within-cell effects on both sides, and with knocked-down cells on both sides; all four lines have random halves only. For the X-Atlas/Orion pair, per-gene random-half response sizes and remaining target expression were compared for genes whose remaining expression differs by at most 0.1, and in bins of the difference; the premise that a gene's remaining expression is reproducible between halves (Spearman ≥ 0.5) was checked first and was not met (0.45); the matched and unmatched ratios were bootstrapped together to give an interval for their difference.

\subsection{Transfer model and benchmark scoring}

A source cell type's total effect is used as the prediction for a target cell type, on shared genes and shared response genes. Ground truth is the target cell type's effect over all cells or over knocked-down cells only, each the mean of two replicates; for hESC both ground truths use guide halves, on the genes where both exist. The source's cross-signal \emph{Q} uses random halves of the source screen and is the same for both ground truths. Denoised quantities use cross-signals, with \emph{S} and \emph{Q} the within-cell-type cross-signals of ground truth and prediction and \emph{X} the inner product of their replicate means: the benchmark's scale factor is √(\emph{S}/\emph{Q}), the relative error (\emph{Q} − 2\emph{X} + \emph{S})/\emph{S} (1 for predicting no change) and the de-attenuated cosine \emph{X}/√(\emph{QS}), each computed per gene and summarised by its median over genes. The single-coefficient transfer model predicts ground truth as β times the source effect; its least-squares coefficient pools genes, β = Σ\emph{X}/Σ\emph{Q}, computed for all-cell and knocked-down ground truth on the same genes (Supplementary Table S5). The noise share of ground truth is 1 − \emph{S}/\emph{T}, with \emph{T} the mean squared norm of the two replicates.

\subsection{Library share of heavy dilution}

In the K562 day-8 and CD4 donor-1 resting genome-scale data and in the essential-enriched hESC library, every recovered perturbation was classed as unmeasurable (fewer than 20 cells), remaining expression not estimable, predicted dilution (1 − \emph{e}\textsubscript{R})\textsuperscript{2} \textless{} 0.5, or none of these. Heavy dilution (the first and third classes) is reported in excess of the non-essential stratum and weighted by stratum size.

\subsection{Decomposition into composition, within-cell change and selection}

With \emph{v}\textsubscript{t,k} the library share of stage \emph{k} among a perturbation's cells and \emph{q}\textsubscript{c,k} the per-UMI rate in stage \emph{k} of control cells, the measured log ratio \emph{E} = log2(\emph{P}\textsubscript{t}/\emph{P}\textsubscript{c}) decomposes exactly as \emph{E} = \emph{E}\textsubscript{c} + \emph{E}\textsubscript{w}, where \emph{E}\textsubscript{c} = log2((Σ\textsubscript{k} \emph{v}\textsubscript{t,k} \emph{q}\textsubscript{c,k})/\emph{P}\textsubscript{c}). Terms become response sizes by the same cross-signal, so \emph{M}\textsubscript{total} = \emph{M}\textsubscript{comp} + \emph{M}\textsubscript{within} + a cross term. With \emph{c} = \emph{M}\textsubscript{comp}/\emph{M}\textsubscript{total} and \emph{w} = \emph{M}\textsubscript{within}/\emph{M}\textsubscript{total} per perturbation, the lower bound of the composition share is \emph{c}, its upper bound 1 − \emph{w}, and the symmetric share (1 + \emph{c} − \emph{w})/2. The selection term is \emph{M}\textsubscript{kd}/\emph{M}\textsubscript{total}. Each run reports the same quantities with the cycle genes removed, together with the null-perturbation control. For the pluripotency-state version in hESC, each cell's score is the mean log1p CP10K of ten pluripotency markers (POU5F1, NANOG, SOX2, LIN28A, DPPA4, TDGF1, L1TD1, DNMT3B, ZFP42, PRDM14) minus that of twelve early differentiation markers (KRT8, KRT18, HAND1, GATA3, GATA6, CDX2, TBXT, MIXL1, PAX6, SOX17, VIM, TFAP2A), cells are binned by the tertiles of control scores, and the three bins take the place of stages.

\subsection{Direction with knocked-down cells}

For hESC genes with at least 10 knocked-down cells in each guide half, ρ with K562 was computed with all cells and with knocked-down cells. The change is reported as the median per-gene difference with a 95\% bootstrap interval; the null drew, in each guide half, the same number of cells as the knocked-down group at random from all of the perturbation's cells, 200 times, and its 2.5th and 97.5th percentiles bound the change expected from cell number alone.

\subsection{Null-perturbation control}

Perturbations whose target is below 0.01 CP10K in a cell type's controls act as null perturbations and should give a response size near zero. They are referenced against the mean of the null perturbations, not the mean of all perturbations.

\subsection{Intervals}

Within-experiment intervals (Supplementary Fig. S12) are 95\% percentile bootstraps over genes (1,000 or 2,000 draws, as listed in Supplementary Table S8). They capture variation between genes; variation between cells and clones enters through the use of independent replicates for response sizes. Replicate values are listed individually wherever more than two biological replicates exist. Design sensitivity is the range over alternative replicate definitions and thresholds. Where a cell type has exactly two independent replicates, no between-replicate interval is claimed. Comparisons across strata and cell types are descriptive and not corrected for multiple testing.

\subsection{Software}

Analyses were run in Python with NumPy, SciPy, h5py and pyarrow. Each analysis script, its output and the number it produced are listed in Supplementary Table S8.

\subsection{Datasets not used, and why}

An activation screen in primary CD4+ T cells\cite{ref24} was excluded because the classification used here is defined for loss of function. A CRISPR interference screen in iPSC-derived microglia\cite{ref7} was excluded because the deposited data have no per-cell guide assignment.

\backmatter
\bmhead{Supplementary information}
Supplementary Figures S1--S12 and Supplementary Tables S1--S8 accompany this preprint, numbered by first citation. Tables: outcomes of knockdown per cell type and stratum (S1), the two-group fits in seven cell types (S2), the hESC--K562 comparison per gene (S3) and across thresholds (S4), benchmark metrics and the transfer model (S5), composition, within-cell change and selection (S6), the genes entering each analysis (S7), and the ledger mapping each number to its script and output (S8); S3, S5, S6 and S8 are data files. Figures: the hESC library (S1), controls for the two-group test (S2), the time course (S3), the classification control and the aggregate estimate (S4), estimators of the hESC--K562 ratio (S5), two further published comparisons (S6), noise share and library share (S7), composition (S8), depletion beyond dependency (S9), and diagnostics of stage calls, per-cell classification and intervals (S10--S12).
\par\noindent\begin{minipage}{\textwidth}
\bmhead{Data availability}
All data analysed here are published. The escape fraction of every fitted perturbation is given in Supplementary Data 1. Accessions, as used in this work:
\par\smallskip
\footnotesize\setlength{\tabcolsep}{3pt}\renewcommand{\arraystretch}{1.15}
\begin{tabularx}{\textwidth}{@{}>{\raggedright\arraybackslash}p{0.34\textwidth}>{\raggedright\arraybackslash}p{9mm}>{\raggedright\arraybackslash}X@{}}
\toprule
Cell type & Source & Accession or location \\
\midrule
K562 (essential, day 6; genome-scale, day 8) and RPE1 (day 7) & ref.~\citenum{ref1} & figshare deposit of the study (files \url{K562_essential_raw_singlecell_01.h5ad}, \url{K562_gwps_raw_singlecell_01.h5ad}, \url{rpe1_raw_singlecell_01.h5ad}) \\
Jurkat and HepG2 (day 7) & ref.~\citenum{ref2} & GEO GSE264667 \\
Primary CD4+ T cells, four donors, resting and stimulated & ref.~\citenum{ref3} & GEO GSE314342; per-cell files also at \url{genome-scale-tcell-perturb-seq.s3.amazonaws.com/marson2025_data/} \\
hESC, genome-scale CRISPRi & ref.~\citenum{ref4} & GEO GSE295214; library composition from Supplementary Table 6 of ref.~\citenum{ref4} \\
iPSC, 19 lines, days 3-5 & ref.~\citenum{ref5} & figshare deposit of the study (guide UMI counts and cell metadata) \\
HCT116 and HEK293T, X-Atlas/Orion (day 7) & ref.~\citenum{ref8} & Hugging Face \url{Xaira-Therapeutics/X-Atlas-Orion}; processed data \url{doi:10.25452/figshare.plus.29190726}; preprocessed gene axis from \url{arcinstitute/PIE_x_atlas_orion} \\
H1 hESC, Virtual Cell Challenge training, validation and test & ref.~\citenum{ref12} & \url{virtualcellchallenge.org/datasets} \\
16 cancer lines, Cas9 knockout, day 6 & ref.~\citenum{ref21} & figshare item 33273600 (\url{processed.zip}, \url{metadata}) \\
K562 time course, CRISPRi and knockout, days 4-14 & ref.~\citenum{ref20} & Zenodo record 20722064 \\
K562 mismatched-guide dose titration (loaded day 5 after transduction) & ref.~\citenum{ref22} & GEO GSE132080 \\
iPSC-derived astrocytes, vehicle and cytokine & ref.~\citenum{ref6} & GEO GSE182308 \\
iPSC-derived microglia (assessed, not used) & ref.~\citenum{ref7} & GEO GSE178317 \\
Primary CD4+ T cells, CRISPRa (assessed, not used) & ref.~\citenum{ref24} & GEO GSE190604 \\
Gene dependency & ref.~\citenum{ref19} & DepMap 24Q4, \url{CRISPRGeneEffect.csv} \\
\bottomrule
\end{tabularx}
\end{minipage}\par
\bmhead{Code availability}
The analysis and figure code is available at https://github.com/xuanruihuang/which-cells-are-measured; an earlier release (1.0.0) is archived at https://doi.org/10.5281/zenodo.23480448, and the scripts added for this version are listed in the repository's README. It includes a function that estimates the escape fraction of each perturbation from per-cell counts. Each number in the text and figures is mapped to the script that produced it and to its output in Supplementary Table S8.
\bmhead{Ethics}
Only published, de-identified data were analysed. The primary CD4+ T-cell data were generated by the original study\textsuperscript{3} from donor material under that study's ethics approvals; no new human material was used.
\bmhead{Competing interests}
The author declares no competing interests.
\bibliography{refs}

\end{document}